\documentclass[10pt, aps,prc,twocolumn,superscriptaddress,preprintnumbers,
amsmath, 
floatfix,
longbibliography,
nofootinbib
]{revtex4-1}
\usepackage[T1]{fontenc}
\usepackage{adjustbox}          % Used to adjust figure frames
\usepackage[caption=false]{subfig}
\usepackage{url}
\usepackage{color}
\usepackage{float}
\usepackage[pdftex,colorlinks=true, linkcolor = blue, citecolor=blue,urlcolor=blue, bookmarksnumbered=true, bookmarksopen=true]{hyperref}
\usepackage{longtable}
\usepackage{amsfonts}
\usepackage{dsfont}
\usepackage{wrapfig,bm} 
\usepackage[normalem]{ulem}
\usepackage{MnSymbol}
\usepackage{float}

\newcommand{\beq}{\begin{equation}}
\newcommand{\eeq}{\end{equation}}
\newcommand{\bea}{\begin{eqnarray}}
\newcommand{\eea}{\end{eqnarray}}
\newcommand{\Tr}{\textrm{Tr}}

\begin{document}
\title{Examining the justification for the introduction of a fermion localization function}
%\title{Is the introduction of a fermion localization function justified? }

\author{Aurel Bulgac}% 
%\email{bulgac@uw.edu}%
\affiliation{Department of Physics,%
  University of Washington, Seattle, Washington 98195--1560, USA}
   
\date{\today}

\begin{abstract}

Becke and Edgecombe suggested in 1990 a theoretical tool to describe electron 
localization in atoms and molecules, 
an idea which was borrowed by a large number of nuclear theorists 
since 2011 to describe nucleon 
localization in nuclear systems. I argue here that these arguments are highly questionable and 
cannot be used in interacting systems, where effects beyond the naive mean field or the simple Hartree-Fock 
framework are important and the inclusion of correlations induced by particle interactions is
necessary in  order to introduce such a localization function. I also describe several aspects of the 
exchange and irreducible 2-body density matrices, which depend on the character and strength 
of the 2-particle interaction and, which can be useful in justifying the derivation of an appropriate 
energy density functional. 
  
\end{abstract}  

\preprint{NT@UW-23-14}

\maketitle   

In nuclei lighter than $^{40}$Ca 
molecular-like states have been studied for a long time~\cite{Morinaga:1956,Funaki:2015}. 
The nuclear molecular
clusters are in reality micro-crystals, as the relative separation between 
the clusters varies very little. In the crust of neutron stars
the formation of the so-called pasta-like phase, another type of matter clusterization,  
or matter crystallization more accurately,
is known for decades~\cite{Ravenhall:1983}. Obviously, the formation of fission 
fragments in nuclear fission~\cite{Meitner:1939} is another 
example of dynamical formation of nuclear clusters. 
In order to easily identify theoretically the formation of ``clumps'' of electronic matter in atoms and 
molecules \textcite{Becke:1990} advocated the use in the Hartree-Fock 
approximation of the quantity  
\begin{align} 
D_{\tau,\sigma}({\bm r}) = \tau_{\tau,\sigma}({\bm r}) 
-\frac{1}{4} \frac{ |{\bm \nabla} {\text n}_{\tau,\sigma}({\bm r})|^2 }{ {\text n}_{\tau,\sigma}({\bm r}) }
-\frac{ |{\bm j}_{\tau,\sigma}({\bm r})|^2 }{{\text n}_{\tau,\sigma} ({\bm r})}, \label{equ:1}
\end{align}  
where the subscripts  $\tau = n, p$ and $\sigma$ stand for 
isospin and spin respectively in case of nuclear systems, and
$\tau_{\tau,\sigma}({\bm r})$, ${\text n}_{\tau,\sigma}({\bm r})$, ${\bm j}_{\tau,\sigma}({\bm r})$ 
are the kinetic energy, nucleon 
and current number densities respectively.  The current density term is required by Galilean 
invariance~\cite{Negele:1972,Engel:1975,Burnus:2005,Son:2006,Bulgac:2007,Gebremariam:2010,Bulgac:2011a}, 
the form used by nuclear theorists~\cite{Reinhard:2011,Zhang:2016,Schuetrumpf:2017,Schuetrumpf:2017a,
Sadhukhan:2017,
Sadhukhan:2022,Sadhukhan:2022a,Scamps:2018,Jerabek:2018,Scamps:2019,Giuliani:2019,Li:2020,
Umar:2021,Matsumoto:2022,Umar:2023,
Li:2023,Ren:2022,Ren:2022a} (this is very likely an incomplete list of references).
Readers will recognize that $D_{\tau,\sigma}({\bm r})$, related to the kinetic energy density without the last term 
was known to \textcite{Weizsacker:1935} (here in the original form, without the current number density)
\begin{align}
\tau_{\tau,\sigma}({\bm r}) = \frac{3}{5}(6\pi^2)^{2/3}{\text n}^{5/3}_{\tau,\sigma}({\bm r}) 
+\frac{1}{4} \frac{ |{\bm \nabla} {\text n}_{\tau,\sigma}({\bm r})|^2 }{ {\text n}_{\tau,\sigma}({\bm r}) }. \label{equ:2}
\end{align}
The accuracy of second term in Eq.~\eqref{equ:2} was questioned 
many times over the years~\cite{DePristo:1987,Jones:1989,Dreizler:1990,Brack:1997}, 
and gradient expansions lead to a pre-factor 1/36 instead of 1/4, 
and currently  Pad\'e approximants and other paramtetrizations are also considered.
There are several reasons for this intense interest in the gradient term, which originates from 
the definition of the 2-body density matrix (see below), since in the Density Functional Theory 
(DFT)~\cite{Dreizler:1990} the exchange and correlation 
energies, arising from interactions, are treated on a equal footing, unlike in the Hartree-Fock approximation, 
which was used in introducing $D_{\tau,\sigma}({\bm r})$ in  Eq.~\eqref{equ:1}.
Eq.~\eqref{equ:1} is derived from the Hartree-Fock approximation of the 2-body number density
for spin pairs with $S=1, S_z=\pm 1, T = 1, T_z = \pm 1$ only, for which Pauli correlation exists 
\begin{align}
&{\text n}_{2}( \xi,\zeta ) = \frac{1}{2} 
\left [    {\text n}_{1} ( \xi,\xi ) {\text n}_{1}(\zeta,\zeta )  - {\text n}_{1} ( \xi,\zeta ) {\text n}_{1}( \zeta,\xi )
                                                                         \right ], \label{equ:BE}
\end{align}
where $n_{1}(\xi,\zeta)=\sum_k \phi_k(\xi) \phi_k^*(\zeta)$ 
is the Hartree-Fock density matrix expressed through the single-particle wave-functions 
$\phi_k(\xi)$, and $\xi =({\bm r}_1,\sigma,\tau),\zeta = ({\bm r}_2,\sigma',\tau')$.
In the limit  ${\bm s}= {\bm r}_1-{\bm r}_2 \rightarrow {\bm 0}$~\cite{Becke:1990,Reinhard:2011,Negele:1972,Engel:1975,Son:2006,
Gebremariam:2010,Bulgac:2011a}
\begin{align}
n_{2}( \xi,\zeta  )= 
\frac{1}{3}{\text n}_{\tau,\sigma}({\bm r}) D_{\tau,\sigma}({\bm r}){\bm s}^2 + {\cal O}({\bm s}^4)\nonumber
\end{align}
and the conditional probability to find a particle with coordinate ${\bm r}'$ from a particle with coordinate ${\bm r}$
and the same spin and isospin is proportional to $D_{\tau,\sigma}({\bm r}) $.
The fermion localization function (FLF) $C_{\tau,\sigma}({\bm r}) $ was introduced in Ref.~\cite{Becke:1990,Reinhard:2011} 
basically as a measured of the accuracy of the zeroth-order Thomas-Fermi approximation for the kinetic energy density
\begin{align}
C_{\tau,\sigma}({\bm r}) = 
\left [ 1+\left (  \frac{D_{\tau,\sigma}({\bm r}) }{\frac{3}{5}(6\pi^2)^{5/3}{\text n}^{2/3}_{\tau,\sigma}({\bm r}) } \right )^2\right ]^{-1}. 
\label{equ:C}
\end{align} 
It appears that in atomic and molecular systems, where the role of correlation energy is significantly 
less important than in nuclei and the Hartree-Fock 
approximation is sufficiently accurate~\cite{Jerabek:2018}, the spatial profiles of the 
electronic shells more pronounced 
in $C_{\sigma}({\bm r})$ than in number density profiles ${\text n}_\sigma({\bm r})$.  
As the term proportional to $|{\bm \nabla} {\text n}_{\tau,\sigma}({\bm r})|^2$ has 
a subdominant role in DFT~\cite{DePristo:1987,Jones:1989,Dreizler:1990,Brack:1997}, 
likely it can be  neglected, 
and in the case of static systems the quantity $C_{\tau,\sigma}({\bm r}) $ is 
defined basically by the ratio of the actual kinetic energy 
density to its Thomas-Fermi approximation. It is obvious that neither this 
ratio nor the quantity $C_{\tau,\sigma}({\bm r}) $
defines an observable and this is merely a somewhat arbitrary measure of the 
accuracy of the Thomas-Fermi approximation 
to the kinetic energy density alone. 

From many studies performed in nuclear physics over the years, it is clear that the 
Extended Thomas-Fermi approximation, including corrections up to ${\cal O}(\hbar^4)$ 
for the kinetic energy density provides a 
pretty good approximation for densities and total energies of nuclear systems, 
see Section 4.4 in Ref.~\cite{Brack:1997}, where 
references to many more complete studies can be found. The (extended) Thomas-Fermi approximation however does 
not describe shell effects, or more generally the quantization of the 
single-particle motion in various geometries. 

Does the quantity $C_{\tau,\sigma}({\bm r}) $ indeed help us better visualize clustering effects?
Cluster formation are typically due to the existence of 2-, 3-, 4-body and higher 
interactions between particles, and not the result of 
the quantization of the single-particle motion in a finite system.    
The main part of the interaction between electrons is repulsive, and in any physical system
(except hard balls at high density) clusters would form only if there will be an effective 
attractive interaction between the electrons. One might argue that the Fock exchange 
Coulomb energy between electrons is attractive, a result which is however accurate only 
for non-relativistic systems~\cite{Dreizler:1990}. However, it is hard to make the argument 
that the ``attractive'' electron Coulomb exchange energy between electrons with the same spin, or in 
spin-polarized electron systems, leads to electron clusterization or electron shells. 
On the other hand, in nuclear systems, which are typically bound and therefore 
the interparticle interaction is mainly attractive, and volume and symmetry energies favor spin- 
and isospin unpolarized systems, the exchange interaction is repulsive, and again, 
one can hardly make the case that the spin- and isospin-polarized nuclear systems can clusterize
or lead to quantization of the single-particle motion. 
One might bring as a counter-example the Wigner crystal~\cite{Wigner:1934} 
of a very low density electron gas, which however is basically
a classical system, where exchange effects are negligible, see also below. The Wigner crystal is 
similar to a system of hard spheres, 
which does not clusterize, but the negligible effects of the exchange 
energy could lead to a disorder state of spins.  Clearly, the quantity $C_{\tau,\sigma}({\bm r}) $, 
which is sensitive to the quantization of 
the single-particle motion and is by definition a 1-body quantity, 
cannot describe the formation of clusters, which 
implies a very strong spatial correlation between various particles, 
and which therefore should be described by a 2-body
or many-body number density. 

In nuclei and in cold fermionic atomic systems, which are qualitatively similar 
to dilute neutron matter~\cite{Zwerger:2011},  
the situation is much more complex. It makes sense to discuss at first the cold 
atom systems, where the interaction is 
very simple and both experimental and {\it ab initio} theoretical approaches are in complete agreement 
to a very high degree of accuracy. 
For a Fermi gas with the zero-range interaction the properties of 
the system are controlled by a single dimensionless 
parameter $k_Fa$, where $a$ is the $s$-wave scattering length, $k_F$ is the 
average Fermi momentum, and where 
there is a complete understanding of both infinite homogeneous systems as well as systems in 
external traps~\cite{Bulgac:2007,Bulgac:2011a}. Around unitarity, 
where the scattering length $|a| \gg 1/\sqrt[3]{n}$ only two fermions 
interact, one with spin up 
and the other with spin down, and in the limit
$|a|\rightarrow +\infty$ the entire system is a gas of barely overlapping pairs with total spin $S=0$, thus a 
clustered system.   In such a Fermi system Cooper 
pairs are formed, with size ranging from extremely large to smaller 
than the average interparticle separation, depending on the actual 
value of $a$, and these Cooper pairs freely collide with each other without being destroyed.  
The 2-body number density for a particle 
with spin-up and the other with spin-down has a universal  behavior  
${\text n}_2({\bm r}_1,{\bm r}_2)\propto 1/|{\bm r}_1-{\bm r}_2|^2$ at any energy or temperature
for $|{\bm r}_1-{\bm r}_2| < |a|$. This behavior is related to a 1-body 
momentum distribution for large momenta $n_p\propto 1/p^4$~\cite{Bulgac:1980,Bulgac:2002,
Bulgac:2002a,Bulgac:2007,Tan:2008a,Tan:2008b,Tan:2008c,Bulgac:2011a,Zwerger:2011,Gandolfi:2011,Carlson:2015,
Bulgac:2019,Bulgac:2022,Bulgac:2022c,Bulgac:2023}.
This system is very similar to the dilute neutron matter appearing in 
neutron stars~\cite{Roggero:2014,Wlazlowski:2014,Tews:2016}. These essentially independent ``Cooper pairs'' are clusters.  

The method suggested by \textcite{Becke:1990} and used in nuclear studies cannot capture 
this kind of clustering in either dilute neutron matter,  nor in 
nuclei~\cite{Reinhard:2011,Zhang:2016,Schuetrumpf:2017,Schuetrumpf:2017a,
Sadhukhan:2017,Sadhukhan:2022,Sadhukhan:2022a,Scamps:2018,Jerabek:2018,
Scamps:2019,Giuliani:2019,Li:2020,Umar:2021,Matsumoto:2022,Umar:2023,
Li:2023,Ren:2022,Ren:2022a}, since 
it considers the correlations or clustering between fermions with the same spin only. 
An {\it ad hoc} alternative was however adopted in nuclei to describe clustering 
or localization on nucleons, and one uses the product ${C_{n,\sigma}({\bm r})C_{p,\sigma}({\bm r})}$. 
It is obvious on the other hand that a product of probabilities describes independent events and therefore 
the proton and neutron distributions cannot be correlated, and therefore this 
measure actually points to the absence 
of clustering. 
The clustering in nuclear systems cannot be explained through exchange effects, but through the 
interplay among volume energy,  surface tension, and symmetry energy,
in a region with a nuclear matter with both protons and neutrons, 
see for classical example the case of the nuclear pasta phase, which actually is a quantum 
crystal~\cite{Ravenhall:1983}, 
as are the nuclear molecular states~\cite{Morinaga:1956,Funaki:2015} as well.
A cluster has a well defined surface, which is characterized by a repulsive surface energy, 
due to nuclear surface tension. Surface tension can be counteracted by a stronger, 
typically proton-neutron attraction, iff proton and neutron densities are (almost) spin saturated.   

The kinetic energy density for fermions with the same spin in a large momentum interval 
exhibits a power law behavior 
$n_p\propto 1/p^4$ in the presence of either pairing correlations and/or
 SRCs~\cite{Bulgac:1980,Bulgac:2002,Bulgac:2002a,
Bulgac:2007,Tan:2008a,Tan:2008b,Tan:2008c,Bulgac:2011a,Bulgac:2019,Bulgac:2022,Bulgac:2022c,Bulgac:2023},
and that would lead to a false signal using Eq.~\eqref{equ:C}.
The SRCs in nuclei are due to several effects, the tensor interaction between protons and 
neutrons~\cite{Levinger:1951,Levinger:2002,Hen:2014,Carlson:2015,
Hen:2017,Cruz-Torres:2018,Cruz-Torres:2021} and pairing  
correlations~\cite{Bulgac:1980,Bulgac:2002,Bulgac:2002a,Bulgac:2007,Tan:2008a,Tan:2008b,Tan:2008c,Gandolfi:2011,
Bulgac:2011a,Zwerger:2011,Bulgac:2019,Bulgac:2022,Bulgac:2023,Bulgac:2022c}, 
the latter also leading to long-range correlations. It is important to 
appreciate that in the case of zero-range interactions the SRCs 
between two fermions, which for a dilute atomic gas are between fermions of the same kind, 
but with opposite spins, are present  at any excitation 
energy~\cite{Tan:2008a,Tan:2008b, Tan:2008c,Zwerger:2011}, 
even when a pairing  condensate, characterized by 
long-range phase order, does not even exist for example in nuclear systems~\cite{Bulgac:2016,
Bulgac:2019c,Bulgac:2020,Bulgac:2020d,Bulgac:2022,Bulgac:2022c,Bulgac:2022d,Magierski:2022}. 

In order to describe such clustering or SRCs using Eq.~\eqref{equ:C} one has to  
consider the  generalization of the FLF
\begin{align}
&C_\tau({\bm r}) = 
\left [ 1+ \left ( \frac{{\cal D}_\tau({\bm r}) }{\frac{3}{5}(3\pi^2)^{2/3}{\text n}_\tau^{2/3}({\bm r})}\right )^2\right ]^{-1}, 
\label{equ:4}\\
&{\cal D}_\tau({\bm r}) = \tau_\tau({\bm r}) - \Delta_\tau ({\bm r}) \kappa_\tau({\bm r})  \nonumber \\
&\quad \quad \quad -\frac{1}{4} \frac{ |{\bm \nabla} {\text n}_\tau({\bm r})|^2 }{ {\text n}_\tau({\bm r}) }
-\sum_\tau \frac{ |{\bm j}_{\tau}({\bm r})|^2 }{{\text n}_\tau ({\bm r})}, \label{equ:5}
\end{align} 
with the very critical important correction arising from the pairing interaction, here for $nn$- or $pp$-pairs only. 
Here ${\text n}_\tau({\bm r}), \, \tau_\tau({\bm r}),\, \kappa_\tau({\bm r})$, and $\Delta_\tau({\bm r})$ 
are the regularized number, kinetic energy, 
anomalous densities, and the renormalized corresponding pairing potentials. 
In the case of zero-range pairing interaction 
the kinetic energy density and anomalous number density both diverge, but the combination
$\tau_\tau({\bm r}) - \Delta_\tau ({\bm r}) \kappa_\tau({\bm r})$ in Eq.~\eqref{equ:4} is free of
divergencies~\cite{Bulgac:2002,Bulgac:2002a,Bulgac:2007,Borycki:2006}. 
Since in nuclei the SRCs are mostly due to the presence of the tensor interaction 
between the protons and neutrons, 
their effect can be described by introducing an effective  $pn$-pairing field~\cite{Bulgac:2022}.
For nuclear systems, as experiments also amply demonstrate~\cite{Levinger:1951,Hen:2014,
Hen:2017,Cruz-Torres:2018,Cruz-Torres:2021} there is always a momentum interval 
where the nucleon momentum distribution has power low behavior $n(p)\propto 1/p^4$ 
as obtained in various {\it ab initio} 
studies~\cite{Carlson:2015,Schiavilla:2007,Bulgac:2022d} and references therein.
The correction due the presence of pairing or/and SRCs has never been considered in discussing nucleon 
localization in previous studies and it is clear that the definition of the 
function ${\cal D}({\bm r})$ then depends on arbitrary  cutoffs used in theoretical 
calculations~\cite{Reinhard:2011,Zhang:2016,Schuetrumpf:2017,Schuetrumpf:2017a,
Sadhukhan:2017,Sadhukhan:2022,Sadhukhan:2022a,Scamps:2018,Jerabek:2018,
Scamps:2019,Giuliani:2019,
Li:2020,Umar:2021,Matsumoto:2022,Umar:2023,Li:2023,Ren:2022,Ren:2022a}.
The use of finite short-range interactions, such as Gogny interaction, 
might superficially mask the presence of divergences, as there is always 
a relatively large momentum interval contributing 
to the kinetic energy and anomalous densities controlled by the short  radius of the 
interaction and where the momentum distribution has a power-law behavior. 

Even after introducing the renormalized quantity ${\cal D}_{\tau}({\bm r})$, 
see Eq.~\eqref{equ:5}, it is still not obvious 
that $C_\tau({\bm r})$ is satisfactory measure of clusterization in nuclear physics, 
since $C_\tau({\bm r})$ is not an 
observable.  As $C_\tau({\bm r})$ is \emph{ a measure of how good 
the Thomas-Fermi approximation only}, and that since in DFT the gradient term of the nucleon density 
$\frac{ |{\bm \nabla} {\text n}_\tau({\bm r}|^2 }{ {\text n}_\tau({\bm r}) }$ can likely be 
neglected~\cite{DePristo:1987,Jones:1989,Dreizler:1990,Brack:1997}, and in the absence of any currents, 
the only thing left in Eq.~\eqref{equ:C} is the ratio of the renormalized kinetic energy density 
to its Thomas Fermi approximation, which again, it is not an observable, unlike the presence of a cluster. 

The tensor interaction plays a very important role in nuclei, 
in particular it leads to a bound proton-neutron system. 
The role of SRCs in nuclei, due mainly to the tensor interaction 
between protons and neutrons has been known for a long time. 
Levinger~\cite{Levinger:1951,Levinger:2002} pointed out more than 70 years ago that SRCs 
are critical to describe the nuclear photo-effect. Since photons are 
a weak probe of nuclear properties, the fact that a pair of neutron and proton 
is predominantly emitted clearly points to the 
presence of short-range quasi-deuteron pairs in nuclei prior to the photon striking a nucleus. 
The presence of strong SRCs in nuclei has been persuasively demonstrated in the JLAB experiment in 
the last decade~\cite{Hen:2014,Hen:2017,Cruz-Torres:2018,Cruz-Torres:2021}.
Obviously, SRCs due predominantly to tensor interaction lead to the formation of clusters, as one can clearly 
see in light nuclei where $\alpha$-like molecular clusters are routinely observed, and 
in which case both proton and neutron subsystems are spin 
saturated and symmetry energy effects dominate, leading to clusters 
with mainly equal proton and neutron numbers. 
A simple evaluation of the magnitude of the symmetry energy, shows that half of its 
magnitude is controlled by minimizing 
the kinetic energy and thus equalizing the spin-up and spin-down occupation 
probabilities of same type of nucleons, 
and the other half is due to the proton-neutron interaction.  As the initial study of localization effects 
in nuclei~\cite{Reinhard:2011} clearly shows, see Fig. 2 in this reference,
the FLF $C_{\tau,\sigma}({\bm r})$  simply very accurately predicts 
where the gradient of the  number density ${\text n}_{\sigma,\tau}$ for a specific $\sigma$ and $\tau$ is largest. 
The volume and symmetry energy ensure that a subsystem is spin- and/or isospin unpolarized, 
unless Coulomb effects become relevant, 
as in the case of the pasta phase in neutron star crust~\cite{Ravenhall:1983}. 
The volume energy and the nuclear surface tension are largely spin and isospin independent.
When an ``internal'' surface appears the single-particle quantization effects 
start playing a subdominant role and they 
appear to be amplified by the FLF ${\cal C}_{\sigma,\tau}({\bm r})$, 
however leading to unrealistic images 
of the nuclear matter distribution. Consider the example of  $^{16}$O discussed in Ref.~\cite{Reinhard:2011}, 
where the FLF $C_{n,\sigma}({\bm r})$ 
allegedly points to the existence of ``spatial shell-like'' structure 
of $^{16}$O with an average  radius of about 3 fm or larger and 
an very pronounced ``inner density depression'' with a radius of about 1.25 fm (estimated at half-density). 
One might surmise that the proton localization should be very similar. At the same time no conceivable 
density probe of $^{16}$O ever revealed the existence of such a pronounced clustering effect, 
specifically the existence of a well-defined spherical shell structure in the number density.
Is there any other type of probe to reveal the reality of this type of clustering? 
The FLF $C_{\tau,\sigma}({\bm r})$ is at best some rather arbitrary measure  
of the accuracy of the Thomas-Fermi approximation to the spin and isospin kinetic energy density, 
and has a very tenuous relation with possible clustering effects, 
which are controlled by the interplay of the surface tension, 
the local spin and isospin saturation in a given nucleus.
In larger nuclei and particularly in neutron start crust  there are significant effects due to Coulomb interaction, 
which only indirectly reflect on the accuracy of the Thomas-Fermi approximation 
of the spin-isospin kinetic energy density ``measured'' by the FLF $C_{\tau,\sigma}({\bm r})$. 

The exact 2-body density matrix ${\text n}_2(\xi,\zeta,\zeta',\xi') $ can be represented as a Hartree-Fock like 
contribution due to the 1-body density matrix ${\text n}_1(\xi,\zeta)$,  plus an irreducible 2-body part 
${\text n}_{corr}(\xi,\zeta,\zeta',\xi')$
\begin{align}
&{\text n}_2(\xi,\zeta,\zeta',\xi') = 
\langle \Phi|\psi^\dagger(\xi)\psi^\dagger(\zeta) \psi(\zeta')\psi(\xi')|\Phi\rangle  \nonumber \\
&= \frac{1}{2}[ {\text n}_1(\xi,\xi')    {\text n}_1(\zeta,\zeta')
                   - {\text n}_1(\xi,\zeta'){\text n}_1(\zeta,\xi') ] \nonumber  \\     
& +  {\text n}_{corr}(\xi,\zeta,\zeta',\xi'), \\                                      
&{\text n}_1(\xi,\zeta) = \langle \Phi|\psi^\dagger(\zeta)\psi(\xi))|\Phi\rangle\nonumber \\
&=\sum_k  n_k \phi_k(\xi)\phi_k^*(\zeta), \quad 0\le  n_k \le 1,\\
&\sumint_\xi{\text n}_1(\xi,\xi) = \sum_k n_k= N,\, \langle \phi_k|\phi_l\rangle = \delta_{kl},\\  
&{\text n}_{ex} (\xi,\zeta,\zeta',\xi')= -\frac{1}{2}{\text n}_1(\xi,\zeta'){\text n}_1(\zeta,\xi'),\\
&\sumint_{\xi,\zeta} {\text n}_2(\xi,\zeta,\zeta,\xi)=\frac{N(N-1)}{2}, \label{equ:N2} \\
&\sumint _{\xi,\zeta} {\text n}_{corr}(\xi,\zeta,\zeta,\xi) = -\frac{1}{2} \sum_k n_k(1- n_k)\le 0,
\label{equ:NN}
\end{align}
where $N$ is the particle number, $n_k$ and $\phi_k(\xi)$ are the canonical occupation probabilities and 
canonical single-particle wave functions~\cite{Bulgac:2022c}, 
also known as natural orbitals~\cite{Lowdin:1955,Lowdin:1956},
and $\xi, \zeta, \xi', \zeta'$ are the particle coordinates $\xi = ({\bm r},\sigma,\tau)$ and so forth. 
The irreducible 2-body part of the 2-body density matrix vanishes only in the case of a pure 
Hartree-Fock wave function, when $n_k =0$ or 1 only, and it is only the second rank of the general 
many-body Born-Bogoliubov-Green-KirkwoodYvon (BBGKY) hierarchy 
of many-body reduced density matrices~\cite{Huang:1987}.  Obviously the solution for the entire 
chain of these different rank density matrices is more complicated than the solution of the corresponding 
many-body Schr\"odinger equation an in practice 
the BBGKY hierarchy of equations is 
truncated in practice at the 1-, 2- or 3-body level at most~\cite{Koltun:1972,Bernheim:1974,
Faessler:1975,Adachi:1989,Dukelsky:1998,Schuck:2016,Cipollone:2013,Carbone:2013,Kas:2019}. 
The trace of the exchange + 2-body irreducible density matrices 
is independent of the particle interaction
\begin{align}
\sumint_{\xi,\zeta} [ {\text n}_{ex}(\xi,\zeta,\zeta,\xi)+{\text n}_{corr}(\xi,\zeta,\zeta,\xi) ]\equiv -\frac{N}{2},\label{eq:xc}
\end{align}
which justifies the Kohn-Sham introduction of the exchange-correlation energy density~\cite{Kohn:1965} 
and which naturally follows form the Kohn-Hohenberg theorem~\cite{Hohenberg:1964}, which states that 
there is a one-to-one correspondence between the many-body wave function and the 1-body density distribution.
There are many methods suggested over the years to go the beyond the Hartree-Fock approximation, and perhaps the most common is the 
shell model in different incarnations, the most recent version being the valence-space 
in-medium similarity renormalization group method (VS-IMSRG)
in nuclear physics, recently extended to atomic systems, see Ref.~\cite{Tenkila:2022} where comparison with other 
approaches, such as coupled clusters and configuration interaction, many-body perturbation theory, 
and earlier references are available. 
IMSRG framework requires however the construction of operators for observables in a reduced space, 
in this case for the density matrix ${\text n}_2(\xi,\zeta,\zeta',\xi')$,
which is not a simple and very transparent procedure~\cite{Tropiano:2022}.
The total, mean field, and correlation energies of a system with 2-body interactions
$V_{2b}$ only are given by
\begin{align}
&E_{tot} = \Tr(T {\text n}_1) + \Tr({\text n}_2V_{2b}),\\
&E_{mf} = \Tr ( {\text n}_{2,mf} V_{2b}), \quad {\text n}_{2,mf} = {\text n}_2 -{\text n}_{corr},\\  
&E_{corr} = \Tr ({\text n}_{corr}V_{2b}),
\end{align}
where $T$ is the kinetic energy and
the trace is an integral over all spatial coordinates and a sum over spin-isospin coordinates.  
The correlation energy is negative for a repulsive $V_{2b}$ interaction and, together with 
the effect of the Fock contribution, leads to a bigger ``Fermi hole,'' particularly in the case of 
a short-ranged repulsive interaction, as one would expect (assuming the
single-particle occupation probabilities $n_k$ do not change).  
The opposite happens in the case of an attractive interaction.
Eq.~\eqref{eq:xc} suggests that the exchange and correlation effects act in opposite directions, a trend partially 
confirmed by microscopic calculations in the case of the homogeneous electron 
gas~\cite{Karasiev:2014,Chachiyo:2016,Karasiev:2016,Dornheim:2018,Kas:2019,Bonitz:2020}. 

 For the sake of the following argument  I introduce
the coupling constant $\lambda$  of the 2-body interaction $\lambda V_{2b}$, 
which is negative for attractive and positive for repulsive particle-particle interaction. 
In general an arbitrary 2-body interaction can have both attractive and repulsive parts, and for simplicity
of the argument I define here a given 2-body interaction $\lambda V_{2b}$ to be attractive if the 
interaction energy $\Tr(\lambda V_{2b}N_{corr})<0$ 
in the limit $\lambda \rightarrow -\infty$ and repulsive otherwise. 
(For interactions with both repulsive and attractive parts this criterion should be applied with care.)
The derivative of the correlation interaction energy with respect to the coupling constant 
(fixed $n_k$, thus first order perturbation in $\delta \lambda$)
\begin{align}
\frac{d E_{corr}}{d \lambda}= \Tr  (V_{2b} {\text n}_{corr})\le 0, 
 \label{equ:n2}
\end{align}
describes the effect of presence of 2-body correlations alone on the 2-particle distributions
in the presence of 2-body interactions beyond the mean field. 
Eq.~\eqref{equ:n2} shows that with increasing strength $\lambda$, from very strong
attractive to very strong repulsive interaction, the correlation energy  
$E_{corr}$ decreases, which in the case of short-range interactions implies
that the ``Fermi hole'' for two identical nucleons becomes bigger.
The extreme values of the trace of 
$\Tr ({\text n}_{corr})$, under the constraint $\sum_kn_k=N$,  are
\begin{align}
-\frac{N}{2} < -\frac{N}{2}\left ( 1- \frac{N}{N_{sp}}\right )  \leq 
\sumint _{\xi,\zeta} {\text n}_{corr}(\xi,\zeta,\zeta,\xi)  \leq 0
\end{align}
which are achieved for 
\begin{align}
&n_k = \frac{N}{N_{sp}}, \quad \text{for \, the \, minimum}\\
& n_k = 0 \quad \text{or} \quad n_k=1\quad \text{for \, the \, maximum} 
\end{align}
where $N_{sp}$ is the number of single-particle states, and which theoretically is infinite. 
The maximum value for $\Tr ({\text n}_{corr})=0$ is 
achieved in the case of a Wigner crystal~\cite{Wigner:1934} 
for long-range very strong repulsive interactions (electron gas)~\cite{Kas:2019,Karasiev:2014,Chachiyo:2016,Karasiev:2016} or in the case of a gas 
for short-ranged very strong repulsive interactions.
The minimum value is attained for a Bardeen-Cooper-Schrieffer superconductor with 
very large attraction and zero spin polarization. In this limit, known as the Bore-Einstein Condensate (BEC) state,
the unpolarized Fermi system is a gas of highly bound dimers/Cooper pairs) with sizes much smaller than the average interparticle separation
and these dimers repel each other~\cite{Randeria:1995,Petrov:2004}. In the limit of a Wigner gas (infinitely repulsive $V_{2b}$) 
$\Tr ({\text n}_{ex})\equiv 0$ and in the opposite BEC limit (infinitely  attractive $V_{2b}$) $\Tr( {\text n}_{ex}) = -N^2/{2N_{sp}}) $ 
is basically vanishing as well.
In the case of attractive short-range interactions 
and finite spin polarization a wide range of phases are 
possible~\cite{Bulgac:2002,Bulgac:2002x,Bulgac:2006,Bulgac:2007a,Bulgac:2008,
Magierski:2009,Magierski:2011,Wlazlowski:2013,Magierski:2019,Bulgac:2020d}.

Since the main argument presented by \textcite{Becke:1990} is based on 
the behavior of the 2-body number density in the limit when 
${\bm r}_1-{\bm r}_2\rightarrow {\bm 0}$ and equal spin and isospins, it becomes obvious that merely the presence of Pauli correlations 
is not a sufficient argument to judge the probability to find another particle nearby in a nuclear medium, in
particularly when they have different spins and/or isospins. Correlations induced by the strong 
particle interactions are crucial and their character depends of whether 
the interaction is attractive or repulsive.
The same rule in Eq.~\eqref{equ:NN} for the 2-body irreducible density matrix 
is expected to be satisfied for any accurate many-body wave function $|\Phi\rangle $
and since in the presence of interactions 
$\sum_kn_k(1-n_k) \neq 0$ it becomes obvious that the procedure suggested
in Ref.~\cite{Becke:1990} and used quite extensively in nuclear 
physics~\cite{Reinhard:2011,Zhang:2016,Schuetrumpf:2017,Schuetrumpf:2017a,Sadhukhan:2017,
Sadhukhan:2022,Sadhukhan:2022a,Scamps:2018,Jerabek:2018,Scamps:2019,Giuliani:2019,Li:2020,
Umar:2021,Matsumoto:2022,Umar:2023,
Li:2023,Ren:2022,Ren:2022a}  cannot describe the clusterization of matter. 
The {\it ad hoc}
procedure adopted in these nuclear studies of the FLF
as ${C_{n,\sigma}({\bm r})C_{p,\sigma}({\bm r})}$
describes independent neutron and proton spin-number densities, 
and ``does not describe correlated neutron-proton subsystems.''
The recent study~\cite{Ren:2022}, pointing to the formation of 
$\alpha$-like structures during nuclear fission is a clear example 
why the FLF is such an inadequate measure, 
as the authors of this study where unable to even determine whether either 
two $^3H$, $^4$He, or $^6$He are present in the neck region.
At the same time proton and neutron density distributions presented in the study~\cite{Ren:2022} fail to show 
the presence of any cluster either before, during or after scission. One can fairly well decide that the
cluster-like structures in the FLFs observed so far in these studies is fictitious or 
simply coincidental at best, since the FLF, 
which is a product  ${C_{n,\sigma}({\bm r})C_{p,\sigma}({\bm r})}$, cannot and does not
describe correlations between proton and neutron subsystems, unlike the irreducible 2-body number density.  

The $nn$- and $pp$-pairing correlations (with $S=0$ and $T=1$) 
alone may lead to a significant correction the FLF.
Similarly, the role of the $np$ SRCs in localization effects is another aspect (summed over isospin), 
which cannot be described with the nuclear FLF
${C_{n,\sigma}({\bm r})C_{p,\sigma}({\bm r})}$, but which can be simulated in a DFT framework by introducing a
dynamic proton-neutron pairing~\cite{Bulgac:2022,Bulgac:2023}.
The presence of pairing correlations leads to another complication, due to the fact that the 
gauge or particle conservation symmetry is broken.In the Hartree-Fock-Bogoliubov (HFB) approximation the 2-body density matrix has the structure
 \begin{align}
 &{\text n}_2(\xi,\zeta,\zeta',\xi') = 
         \langle \Phi_{HFB}|\psi^\dagger(\xi)\psi^\dagger(\zeta) \psi(\zeta')\psi(\xi')| \Phi_{HFB}\rangle 
  \nonumber \\
 &= \frac{1}{2}[ {\text n}_1(\xi,\xi')    {\text n}_1(\zeta,\zeta')
                   - {\text n}_1(\xi,\zeta'){\text n}_1(\zeta,\xi') ] \nonumber  \\     
 & +  \frac{1}{2} \kappa(\xi,\zeta)\kappa^*(\xi',\zeta'),
 \end{align}
 where $\kappa(\xi,\zeta) = \langle \Phi_{HFB}|\psi(\zeta)\psi(\xi)|\Phi_{HFB}\rangle$ is 
 the anomalous density-matrix and the corresponding irreducible 2-body density matrix satisfies now 
 the incorrect condition
\begin{align}
\frac{1}{2}\sumint_{\xi,\zeta} |\kappa(\xi,\zeta)|^2 \ge 0,
\end{align}
with an the opposite sign to Eq.~\eqref{equ:NN} and  as a result the corresponding 2-body 
density matrix does not satisfy anymore the expected sum rule 
defined in Eq.~\eqref{equ:N2}. This aspect can 
be corrected only after the particle projection of the HFB many-wave function $\Phi_{HFB}$ is performed.
Consequently, the improved ${\cal D}_\tau({\bm r})$  introduced in Eq.~\eqref{equ:5} cannot be expected 
to lead to a correct outcome in the presence of pairing correlations, unless particle projection is performed
when evaluating the 2-body density matrix.  The particle projected occupation probabilities $n_k$ and the 
corresponding particle projected HFB 2-body density matrix ${\text n}_2(\xi,\zeta,\zeta',\xi')$ can be easily 
evaluated~\cite{Bulgac:2021}. Pairing correlations are particular example when the  irreducible
density matrix ${\text n}_{corr}(\xi,\zeta,\zeta',\xi')$ plays a large role and the role of exchange effects is reduced.

In summary, I have shown that the use of the FLF introduced by \textcite{Becke:1990} and widely
used in theoretical nuclear studies~\cite{Reinhard:2011,Zhang:2016,Schuetrumpf:2017,Schuetrumpf:2017a,
Sadhukhan:2017,
Sadhukhan:2022,Sadhukhan:2022a,Scamps:2018,Jerabek:2018,Scamps:2019,Giuliani:2019,Li:2020,
Umar:2021,Matsumoto:2022,Umar:2023,
Li:2023,Ren:2022,Ren:2022a} is ill-justified and cannot correctly describe clusterization effects, 
which require the knowledge of the irreducible 2-body number density ${\text n}_{corr}$ 
if a similar approach is adopted. 
The cases of the Wigner crystal and of the unitary Fermi gas in the BEC limit discussed above 
are clear examples where FLF fails to disentangle the ``clusters.'' 
Another example would be a system of two dimers, one a relatively strongly bound ``$S =1$ neutron dimer'' 
and the other a relatively  strongly bound ``$S=1$ proton dimer,'' 
when either these two dimers repel in a trap or are weakly bound. 
Particularly large errors can be incurred when including pairing correlations within the energy density functional, 
which are ubiquitous  nuclear systems discussed so far in literature, and which have a significant 
impact on various nuclear properties. 
 The homogeneous 
electron gas~\cite{Karasiev:2014,Chachiyo:2016,Karasiev:2016,Dornheim:2018,Kas:2019,Bonitz:2020}, 
neutron matter with chiral effective 2- and 3-body interactions~\cite{Roggero:2014,Wlazlowski:2014,Tews:2016} 
and the dilute Fermi gas, particularly in the unitary and BEC regime, are examples of microscopic well-studied
systems~\cite{Randeria:1995,Bulgac:2002,Bulgac:2002x,Petrov:2004,Bulgac:2006,Bulgac:2007,
Bulgac:2007a,Bulgac:2008,Magierski:2009,Magierski:2011,
Bulgac:2011a,Wlazlowski:2013,Magierski:2019,Bulgac:2020d,Bulgac:2011a,Zwerger:2011},
where the role of both exchange, correlations , and temperature effects are important. 
The case of quarks localized inside hadrons is the most notable example 
of  the dominant role of strong correlations effects.

\vspace{0.5cm}
        
{\bf Acknowledgements} \\

I thank G. Scamps and M. Kafker for reading the paper
and asking for clarifications on an initial version, which improved the present version. 
The funding from the US DOE, Office of Science, Grant No. DE-FG02-97ER41014 and
also the support provided in part by NNSA cooperative Agreement
DE-NA0003841 is greatly appreciated.

%%%%%%%%%%%%%%%%%%%%%%%%%%%%%%%%%%%%%%%%%%%%%

% These are needed to avoid a babel error.
\providecommand{\selectlanguage}[1]{}
\renewcommand{\selectlanguage}[1]{}

\bibliography{localization.bib}

\end{document}